\documentclass{article}

\usepackage{arxiv}

\usepackage[utf8]{inputenc} 
\usepackage[T1]{fontenc}    
\usepackage{hyperref}       
\usepackage{url}            
\usepackage{booktabs}       
\usepackage{amsfonts}       
\usepackage{nicefrac}       
\usepackage{microtype}      
\usepackage{lipsum}		
\usepackage{graphicx}
\usepackage[super]{natbib}
\usepackage{doi}
\usepackage{threeparttablex}
\usepackage{sistyle}
\usepackage{url}
\usepackage{amsmath}
\SIthousandsep{,}
\newcommand{\Xtilde}{\widetilde{X}}

\newcommand{\bZ}{\pmb{Z}}
\newcommand{\bS}{\pmb{S}}
\newcommand{\bM}{\pmb{M}}
\newcommand{\bB}{\pmb{B}}
\newcommand{\bbeta}{\pmb{\beta}}

\usepackage{import}
\newcommand{\rev}[1]{#1}

\usepackage{dsfont}

\title{Prediction Modeling from an EHR-Derived Allostatic Load Index with Informatively Missing Biomarkers: A Case Study}

\author{
\href{https://orcid.org/0009-0002-1247-8688}{\includegraphics[scale=0.06]{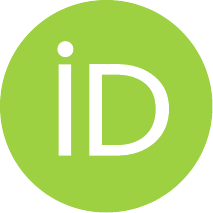}Grayson E. ~Weavil} \\
Department of Statistical Sciences, Wake Forest University, Winston-Salem, North Carolina, U.S.A. \\
Department of Public Health Sciences, Medical University of South Carolina, Charleston, South Carolina, U.S.A. \\
\And 
\href{https://orcid.org/0000-0001-6265-0752}{\includegraphics[scale=0.06]{orcid.pdf}\hspace{1mm}Joseph ~Rigdon} \\ 
Department of Biostatistics and Data Science\\
Wake Forest University School of Medicine \\
Winston-Salem, NC 27157 \\
\And
\href{https://orcid.org/0000-0001-5380-2427}{\includegraphics[scale=0.06]{orcid.pdf}\hspace{1mm}Sarah C.~Lotspeich} \\
	Department of Statistical Sciences, Wake Forest University, Winston-Salem, North Carolina, U.S.A. \\
	\texttt{lotspes@wfu.edu} \\
}

\renewcommand{\shorttitle}{Prediction from EHR-Derived Allostatic Load Index with Informatively Missing Biomarkers}

\hypersetup{
pdftitle={MissALI},
pdfsubject={q-bio.NC, q-bio.QM},
pdfauthor={Grayson ~Weavil, Joseph ~Rigdon, Sarah C.~Lotspeich},
pdfkeywords={},
}

\begin{document}
\maketitle

\begin{abstract}
Embedding a standardized measure of physiologic stress in electronic health records (EHR) could be instrumental to preventative care. The allostatic load index (ALI), calculated from ten biomarkers across three body systems, offers a promising snapshot of cumulative strain. The ALI can be calculated from EHR data, but many components are missing, since not all patients undergo all tests. Using statistical modeling and machine learning, EHR data for $707$ patients from a large academic health system ($17\%$ of whom were hospitalized), were used to predict in-patient hospitalization from ALI, controlling for age and sex. Various methods were evaluated to fill in information gaps for patients' missing ALI components, including summary measures combining components or using them separately. Performance was measured using various discrimination, calibration, and hybrid metrics. 
Discriminative performance was fair to low and calibration low to moderate for both model types. Brier scores indicated good overall accuracy across models. Summary measure approaches performed similarly with respect to most metrics. Separate component methods, especially pattern submodels, showed potential improvement over summary measures in the full sample, but their advantages did not persist under cross-validation. Most summary measure models had moderate discrimination and calibration ability. When cross-validated, separate component models appeared overfit and performed similarly to (or worse than) summary measures. Researchers may want to prioritize whichever summary measure was most interpretable or appropriate for their beliefs about missingness. Next steps include developing refined prediction models from  EHR-derived ALI to improve performance and enhance clinical relevance.
\end{abstract}


\section*{Background}

With rising medical care costs and increased access to patient risk identification tools, there is a push to transition from reactive to proactive healthcare. Learning health systems aim to establish a continuous feedback loop between \emph{practice} (clinicians treating patients), \emph{data} (information collected through patient care), and \emph{evidence} (results of data-based studies), where biomedical research is the driving force between them.\cite{enticott2021} Their goal is to identify at-risk patients before a crisis occurs. 

Electronic health records (EHR) provide a wealth of patient data that can inform clinician decisions and advance the learning health system. However, EHR data are subject to high rates of missing values, because data collection depends on clinical decision-making (e.g., test ordered) and access barriers (e.g., insurance approval). Still, adjusting for missingness, EHR data can be valuable to research, offering large sample sizes and many variables.  

\rev{Allostasis is the body's ability to rebound in response to stress,\cite{STERLING20125} while allostatic load is the cumulative physiological burden of stress on the body.\cite{McEwenStellar1993}} Implementing a measure of allostatic load in the learning health system would enable large-scale monitoring of patient well-being and prediction of key outcomes, as with other scores like the electronic frailty index (EFI) for elderly populations.\cite{Pajewski2019, Orkaby2024} \rev{Several systematic reviews outline drivers of allostatic load and its downstream health outcomes.\cite{Guidi2021} For example, socioeconomic status\cite{Gustafsson2011}, perceived neighborhood quality,\cite{Carbone2020} and adverse childhood events\cite{Danese2012,Dich2015} have all been associated with higher allostatic load. Further, higher allostatic load has been associated with various acute and chronic conditions, including cardiovascular disease,\cite{Seeman1997,Evans2025} mental health disorders,\cite{Juster2011} migraines,\cite{Borsook2012} all-cause and cause-specific mortality.\cite{Evans2025, BruunRasmussen2022,Parker2022} 
The allostatic load has also played a key role in quantifying health disparities.\cite{Geronimus2006, Rodriquez2019, Shalowitz2019, Ketheesan2020_disparity, Harris2022, Borrell2010} 
}

\rev{Since allostatic load is a latent construct (i.e., it cannot be measured directly), computable phenotypes are used to capture it based on observable measures, like biomarkers. There are many such formulations for allostatic load, as seen in the literature cited thus far and detailed in systematic reviews on its measurement.\cite{Mauss2015,Duong2017,Beese2022,Rodriquez2019,Guidi2021, Juster2010AllostaticLoad,Ketheesan2020,Ribeiro2018}}
We focus on the allostatic load index (ALI), measured from ten biomarkers across three body systems (Table~\ref{tab:ali_components}). Traditionally, ALI is the count of unhealthy biomarkers, where the ``unhealthy'' classification is often study-specific.\cite{Mauss2015,Beese2022,Ketheesan2020,Duong2017,Ribeiro2018} \rev{While some studies used data-driven risk quartiles,\cite{Geronimus2006,Seeman1997,Seeman2001,Wallace2013,Juster2011,BruunRasmussen2022} we and others\cite{Merkin2009,Borrell2010,Frei2015,Rosenberg2015,Shalowitz2019} pre-specified thresholds based on standard clinical guidelines. Some have used published clinical thresholds when available and supplemented with quartiles when needed.\cite{Carbone2020,Dich2015}} 

An ALI of zero represents \rev{least stress (most healthy) and ten represents most stress (most unhealthy)}. By capturing multi-system dysregulation, rather than isolated abnormalities, ALI reflects the cumulative effects of chronic stress and biological ``wear and tear,'' 
making it a valuable clinical predictor. \rev{Notably, six of our ALI components were among those most frequently cited in a recent systematic review of reviews on allostatic load measurement, which contained information from $238$ manuscripts.\cite{Beese2022} In particular, hemoglobin A1C, systolic blood pressure (SBP), and diastolic blood pressure (DBP) were included in $>60\%$ of studies.} 

\begin{table}[ht!]
\centering
\begin{tabular}{lr}
        \textbf{Component}& \textbf{Unhealthy Threshold}\\
        \hline 
        \multicolumn{2}{l}{\textit{Cardiovascular System}} \\
        Systolic Blood Pressure 
        & $>140$ mmHG  \\
        Diastolic Blood Pressure 
        & $>90$ mmHG \\
        \multicolumn{2}{l}{\textit{Metabolic System}} \\
        Body Mass Index 
        & $>30$ kg/m$^2$ \\
        Triglycerides 
        & $\geq 150$ mg/dL\\
        Total Cholesterol 
        & $\geq 200$ mg/dL \\
        \multicolumn{2}{l}{\textit{Inflammation System}} \\
        C-Reactive Protein 
        & $\geq 10$ mg/L \\
        Hemoglobin A1C 
        & $\geq 6.5\%$\\
        Serum Albumin 
        & \rev{$\geq 3.5$} g/dL \\
        Creatinine Clearance 
        & $<110$ mL/min(Males) \\
        & $<100$ mL/min (Females) \\
        Homocysteine 
        & $>50$ mcmol/L \\
    \end{tabular}
\caption{Ten biomarkers, spanning the cardiovascular, metabolic, and inflammatory systems, that compose the allostatic load index (ALI) were transformed into binary indicators using \rev{clinical} thresholds. Based on the cutoffs, values were coded as $1$ (unhealthy), and values within normal reference ranges were coded as $0$ (healthy). Abbreviations: Millimeters of mercury (mmHG), milligrams per deciliter (mg/dL), kilograms per square meter (kg/m$^2$), milligrams per liter (mg/L), grams per deciliter (g/dL), milliliters per minute (mL/min), micromoles per liter (mcmol/L)}
\label{tab:ali_components}
\end{table}

Our ALI is based on that of Seeman et al. (1997, 2001),\cite{Seeman1997, Seeman2001} which was first developed in the MacArthur Successful Aging Study, where researchers could ensure that patients underwent all testing for the ten biomarkers. In observational settings, like EHR, there are no such guarantees. Measurements of some ALI components are embedded in routine clinical care (e.g., SBP, DBP, and body mass index [BMI]), making them well-populated in EHR data. However, many other components (e.g., homocysteine and creatinine clearance) are collected only when clinically necessary, making them informatively missing.

Thus, while an EHR-derived ALI can convey patient risks to clinicians and inform institutional planning, missing data remains a challenge to its adoption. For instance, predicting patient engagement in the healthcare system from ALI could help improve patient care (saving lives) and prevent future hospitalization (reducing costs). Yet, calculating ALI without carefully accounting for missingness can lead to inaccurate conclusions. 
\rev{In this case study, we evaluate strategies to handle missing ALI components in an illustrative example predicting in-patient hospitalization with statistical modeling and machine learning techniques.} 

\section*{Methods}

\subsection*{Sampling and data extraction}\label{subsec:cohort}

This study builds upon an EHR-derived sample of $1000$ patients who routinely engaged in care 
at Atrium Health Wake Forest Baptist Hospital (AHWFB) in Winston-Salem, North Carolina between March $11$, $2018$, and March $10$, $2020$. To be eligible \rev{for that sample}, patients had to (i) be $18-65$ years old and (ii) initiate their first primary care outpatient encounter (PCOE) at AHWFB during the $2$-year study period. 
The Institutional Review Board at Wake Forest University School of Medicine approved this study. 

Demographics, vitals, labs, and hospitalizations were extracted for all patients. The outcome of interest was in-patient hospitalization at AHWFB during the $2$-year study period, $Y$. 
ALI component measurements (Table~\ref{tab:ali_components}) were taken from vitals and labs conducted either at AHWFB (before March 10, 2020) or a previous institution $\leq 3$ years prior \rev{to their first PCOE at AHWFB}. Given that most patients contributed multiple encounters, ALI components were defined from mean measurements. Patient age (in years) at their first AHWFB primary care outpatient encounter $Z_1$ (hereafter, ``age'') and biological sex $Z_2$ (male/female) were included as covariates. 

\rev{For our analysis, we imposed two additional criteria. First, to ensure that all patients had sufficient opportunity for hospitalization, we excluded $292$ patients whose first PCOEs were $<90$ days before study end. Second, only labs and vitals $\geq 1$ day before first hospitalization were included in calculating ALI components (where applicable). One patient was excluded who only had labs and vitals from their day of hospitalization. The final dataset included $n = 707$ patients.} 

\subsection*{Allostatic load index (ALI)}\label{subsec:ali}

Our ALI $X$ is calculated from a fixed set of ten components. Let $\bB = (B_1, \dots, B_{10})$ denote the ten biomarker measurements of interest and $\bS = S_1, \dots, S_{10}$ be the corresponding discretized ALI components. Each component $S_j = 1$ if $B_j$ is  unhealthy based on \rev{the threshold} and $S_j = 0$ otherwise ($j \in \{1,\ldots,10\}$).  For example, if $B_{1}$ denotes a patient’s average SBP, then $S_{1}=1$ if $B_1 > 140$ millimeters of mercury and $S_{1}=0$ otherwise.

Then, ALI is traditionally computed as $X = \sum_{j=1}^{10} S_j$,\cite{Seeman1997,Seeman2001} representing the count of unhealthy measurements. An ALI $=0$ reflects optimal \rev{cumulative stress (none)}, whereas an ALI $=10$ indicates the highest possible stress. This definition fails to accurately capture \rev{stress} from incomplete data. Missing values are implicitly replaced with zeros, misclassifying missing components as ``healthy'' and potentially misrepresenting \rev{cumulative physiological stress}.

\subsection*{Modeling hospitalization} \label{subsec:miss_data}

Hospitalization was predicted using two modeling paradigms: statistical regression and machine learning. \rev{Both are popular approaches,  differing in their flexibility and interpretability. We were interested in within-paradigm comparisons of the various missing data approaches for ALI. In other words, after choosing between statistical modeling and machine learning, how can we best account for informatively missing ALI components?}  

\subsubsection*{\rev{Logistic regression}} \label{subsec:stat_mods}

Let $Y=1$ indicate a patient being hospitalized, and $Y = 0$ otherwise, and $\bZ = (Z_1, Z_2)$ represent the vector of covariates. 
For statistical modeling, we used  \textit{logistic regression}: 
\begin{equation}
 \log\left\{\frac{{\Pr_{\bbeta}}(Y = 1 \mid X, \bZ)}{1-{\Pr_{\bbeta}}(Y = 1 \mid X, \bZ)}\right\} = \beta_0 + \beta_1 X + \bbeta_2^{\top}\bZ,  \label{def:log_reg} 
\end{equation}
where the left-hand side of \eqref{def:log_reg} denotes a patient's conditional log odds of hospitalization. 
In R, the built-in \texttt{glm} function fits logistic regression.\cite{R}

\subsubsection*{\rev{Random forest}}\label{subsec:mach_learn}

Machine learning methods can offer greater flexibility in modeling complex, nonlinear relationships. \rev{For illustration, we} focus on random forest. 
We used the \textit{ranger} package in R with default settings.\cite{ranger} \rev{Each of the forest's $500$ ``trees'' was constructed using a bootstrap re-sample of the data. At each split, a random subset of $\sqrt{p}$ predictors was considered, where $p$ is the total number of predictors in the fitted model. The split with the largest reduction in Gini impurity was chosen. This process allowed for variability across trees, and predicted probabilities were aggregated (i.e., ensembled) by averaging across trees to produce a final prediction for each patient.} 

\subsection*{Handling missing ALI components} \label{subsec:miss_data}

Data in the EHR are often missing not at random (MNAR). For example, many inflammation components of ALI depend on labs that are typically ordered based on a clinical need, making \rev{routine} values more likely to be missing.\citep{Haneuse2021} Therefore, missingness itself carries meaningful information about patient health status and hospitalization risk. Still, which is the best way to account for missing components of a score like the ALI \rev{in a prediction model} is an important open question. 

Using logistic regression for demonstration, we primarily considered methods for handling missing data in an aggregate ALI score. These ``summary measure'' approaches retained the concise nature and interpretability of the original ALI, which was practically important. However, we also considered models where ALI components functioned as separate predictors, which could offer improved performance by allowing individual \rev{biomarkers} to impact hospitalization differently. For \rev{random forest}, the methods were adapted in parallel.

\subsubsection*{Complete-case proportion}\label{subsec:app1}

Converting the ALI from a sum to a proportion and excluding missing components (i.e., ignoring them rather than treating them as healthy or unhealthy) is one straightforward modification. The so-called \textit{complete-case proportion} ALI is defined: 
\begin{align}
\Xtilde &= \frac{\sum_{j=1}^{10} R_j S_j}{\sum_{j=1}^{10} R_j} \nonumber \\
&= \frac{\text{Number of non-missing and unhealthy components}}{\text{Number of non-missing components}}
\label{eq:prop_ali}
\end{align}
where $R_1, \dots, R_{10}$ are indicators of components $S_1, \dots, S_{10}$, respectively, being non-missing. Then, $\Xtilde$ can be interpreted as the proportion of a patient's non-missing ALI components that were unhealthy. The minimum $\Xtilde = 0$ corresponds to the best \rev{status} \rev{(all non-missing biomarkers were healthy)}, while the maximum $\Xtilde = 1$ corresponds to the worst (\rev{all non-missing biomarkers were unhealthy}). 

Using $\Xtilde$ in place of $X$ in \eqref{def:log_reg}, the logistic regression model using the complete-case proportion ALI is given by
\begin{equation}
 \log\left\{\frac{{\Pr_{\tilde{\bbeta}}}(Y = 1 \mid \Xtilde, \bZ)}{1-{\Pr_{\tilde{\bbeta}}}(Y = 1 \mid \Xtilde, \bZ)}\right\} = \tilde{\beta}_0 + \tilde{\beta}_1 \Xtilde + \tilde{\bbeta}_2^{\top}\bZ. \label{eq:prop_ali_mod}
\end{equation}
This approach was selected for simplicity and consistency with prior work.\cite{Pajewski2019, lotspeich2025} However, we wondered whether incorporating, rather than ignoring, missing components might improve prediction. 

\subsubsection*{Counts of unhealthy and missing components}\label{subsec:app2}

We next considered an ad-hoc strategy where the original count ALI  was paired with the count of missing components. Let $C_S = \sum_{j=1}^{10} R_{j}S_{j}$ denote a patient's number of non-missing, unhealthy ALI components and $C_M = \sum_{j=1}^{10}(1-R_{j})$ denote their number of missing ALI components. 

Then, using ($C_S$, $C_M$) instead of $X$ in \eqref{def:log_reg}, the logistic regression model using \textit{counts of unhealthy and missing components} is 
\begin{align}
& \log\left\{\frac{{\Pr}_{\bbeta^C}(Y = 1 \mid C_S, C_M, \bZ)}{1 - {\Pr}_{\bbeta^C}(Y = 1 \mid C_S, C_M, \bZ)}\right\} \nonumber \\
&= \beta^C_0 + \beta^C_1 C_S + \beta^C_2 C_M + \bbeta_3^{C\top}\bZ.
\label{eq:num_miss}
\end{align}
This formulation was motivated by the hypothesis that the amount of missingness may be informative about overall well-being (e.g., healthier individuals may undergo fewer laboratory tests). 

\subsubsection*{Best/worst case imputation}\label{subsec:app4}

To assess the potential impact of missing data under extreme scenarios, we implemented simplified imputation strategies corresponding to best- and worst-case scenarios for what each patient's ALI could be if all of their biomarkers had been available. Specifically, we considered \textit{best case imputation}, treating all missing ALI components as healthy, and \textit{worst case imputation}, treating all missing components as unhealthy.

Let the best-case scenario ALI be
\begin{align}
\breve{X}^B &= \frac{\sum_{j=1}^{10}R_jS_j}{10} \nonumber \\
&= \frac{\text{Number of unhealthy components}}{\text{Total number of components}}, 
\label{eq:bc_ali} 
\end{align}
and worst-case scenario ALI be 
\begin{align}
\breve{X}^W &= \frac{\sum_{j=1}^{10}R_jS_j+\sum_{j=1}^{10}(1-R_j)}{10} \nonumber \\
&= \frac{\text{Number of unhealthy or missing components}}{\text{Total number of components}}.
\label{eq:wc_ali}
\end{align}
Then, we estimated \eqref{def:log_reg} using $\breve{X}^B$ or $\breve{X}^W$ in place of $X$. 

These simplified imputation approaches provide lower and upper bounds on \eqref{eq:prop_ali_mod}, since $\breve{X}^B \leq \Xtilde \leq \breve{X}^W$.  Case imputation has been used in clinical trials with missing outcomes,\cite{Gould1980, Higgins2008} where performance depended on how well missingness alone could predict the variable of interest. 
Here, best case imputation was motivated by informative missingness in EHR data, where some biomarkers are only ordered when clinically indicated. Worst case imputation offered a more conservative contrast to this low-risk summary measure. We also considered best/worst case imputation when ALI components functioned separately in models (Supplementary Materials). 

\subsubsection*{Missingness as a category}

There could also be value in allowing patients who are missing particular components to constitute their own category, rather than forcing them into either healthy or unhealthy (as with imputation). For discrete variables, this approach can be seen as a special case of missingness indicators\cite{Groenwold1265} and pattern mixture models.\cite{Rosenbaum1984,Dagostino2000} There is no clear way to collapse information across components, making it not a fair comparator to our summary measure  approaches, but we were interested in quantifying how much information is potentially lost by aggregating the ALI components. 

For each component, let $S_j$ denote whether the component is non-missing and unhealthy and let $M_j = 1 - R_j$ denote whether the component is missing ($j \in \{1, \dots, 10\}$). (If $S_j = 0$ and $M_j = 0$, then component $j$ is non-missing and healthy.) Using these predictors instead of $X$ in \eqref{def:log_reg} the logistic regression model using the \textit{missingness as a category} approach is defined as:
\begin{align}
& \log\left\{\frac{\Pr_{\dot{\bbeta}}(Y = 1 \mid \bS, \bM, \bZ)}{1 - \Pr_{\dot{\bbeta}}(Y = 1 \mid \bS, \bM, \bZ)}\right\} \nonumber \\
&= \dot{\beta}_0 + \sum_{j=1}^{10} \dot{\beta}_{1j} S_j + \sum_{j=1}^{10} \dot{\beta}_{2j} M_j + \dot{\bbeta}_{3}^{\top}\bZ,
\label{eq:miss_ind}
\end{align}
where $\bS$ and $\bM$ denote the vectors of ALI components and missingness indicators, respectively. In our application, some components did not need all three categories (e.g., SBP and DBP which had no missing values). Further, because all patients who were missing triglycerides were also missing cholesterol (they come from the same blood test), there was collinearity. We 
excluded one of their missingness indicators, as is the default in R's \texttt{glm} function. 

This approach challenged the implicit assumption, apparent in summary-based measures, that all ALI components contribute equally for health status and missingness. 
In fact, it generalizes \textit{counts of unhealthy and missing components} (Supplementary Materials). Previous work cautioned about the use of missingness indicators in prediction models when missingness depends on the outcome,\cite{Sisk2023} but we expect missingness in ALI components to depend on the underlying biomarkers, not hospitalization.  

\subsubsection*{Pattern submodels}

\textit{Pattern submodels} were proposed as an alternative to imputation, particularly for MNAR settings. This approach is similar to pattern mixture models, but more tailored to prediction over inference.\cite{Fletcher2020} The basic idea is to take advantage of all unique missing data patterns and fit the best ``submodel'' specifically tailored to patients with that pattern. All other techniques discussed herein fit a single model with all patients. 

\begin{figure}[h]
    \centering
    \includegraphics[width=0.9\linewidth]{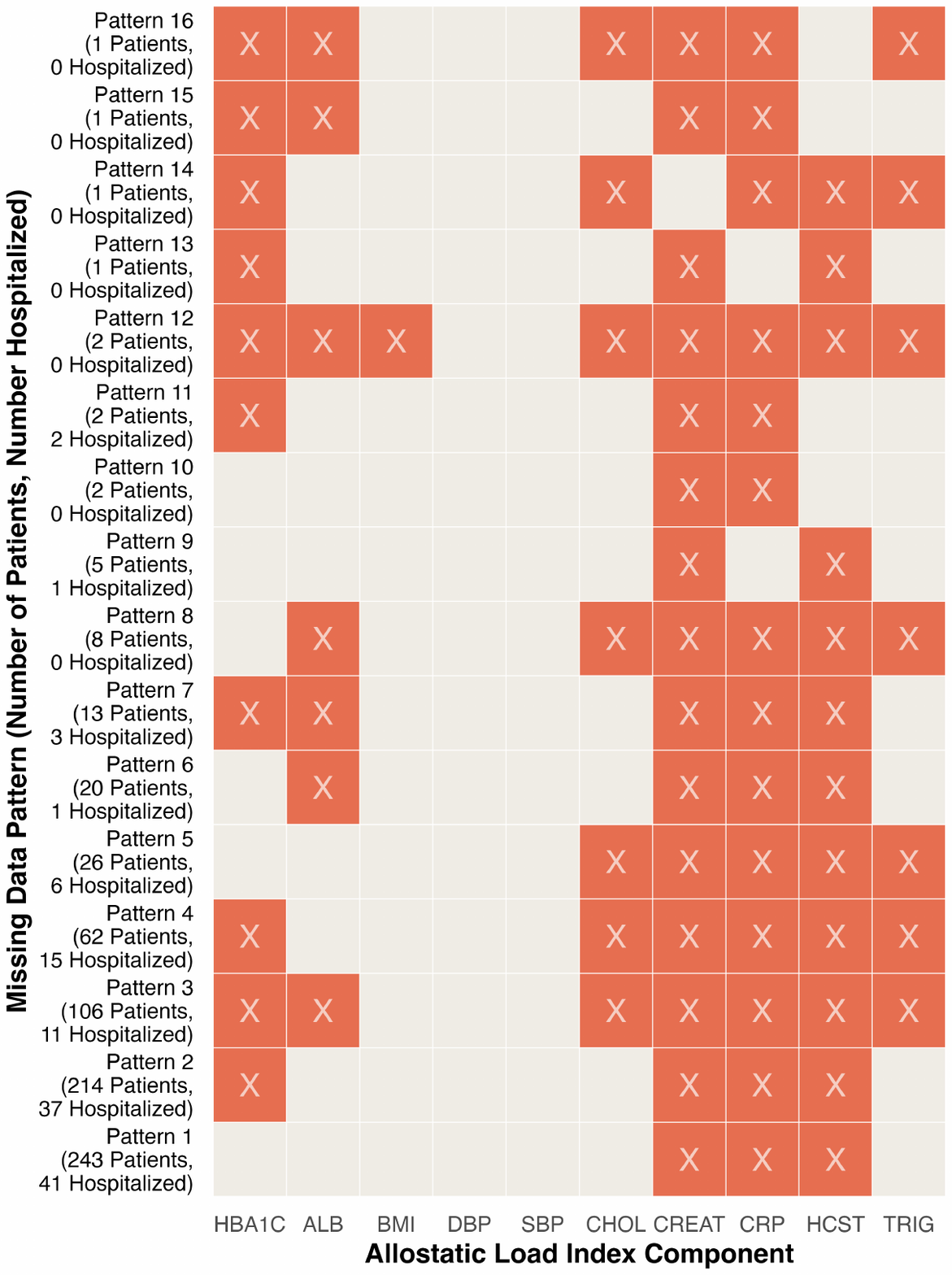}
    \caption{Patterns of missingness in the ten allostatic load index (ALI) components for the $n = 707$ patients in the sample from the electronic health records (EHR) at Atrium Health Wake Forest Baptist Hospital. Each row represents one of the $16$ distinct missing data patterns, and within that row the ALI component(s) with an ``X'' were missing. Abbreviations: hemoglobin A1C (HBA1C), serum albumin (ALB), body mass index (BMI), diastolic blood pressure (DBP), systolic blood pressure (SBP), cholesterol (CHOL), creatinine clearance (CREAT), C-reactive protein (CRP), homocysteine (HCST), and triglycerides (TRIG).}
    \label{fig:miss.pat}
\end{figure}

In our application, there were $K = 16$ unique missing data patterns (i.e., combinations of ALI components missing together), with as few as $1$ and as many as $243$ patients each (Figure~\ref{fig:miss.pat}). 
For each missing data pattern $k$ ($k \in \{1,\dots,K\}$), the logistic regression submodel is 
\begin{align}
& \log\left\{\frac{\Pr_{\Acute{\bbeta}^k}(Y^k = 1 \mid \bS^k, \bZ^k)}{1 - \Pr_{\Acute{\bbeta}^k}(Y^k = 1 \mid \bS^k, \bZ^k)}\right\} \nonumber \\
&= \Acute{\beta}^k_0 + \sum_{j:R_j^k=1}
\Acute{\beta}^k_{1j} S^k_{j} + \Acute{\bbeta}_{2}^{k\top}\bZ,
\label{eq:submod}
\end{align}
where the superscript $k$ corresponds to the pattern and the summation is over all components $j$ that were non-missing for it. When evaluating performance, patients' outcomes were predicted from the appropriate submodel. More details can be found in the Supplementary Materials, including how to handle rare missing data patterns with few patients. 

\subsection*{Evaluating model performance}\label{evaluation}

To evaluate models' performance, we calculated \rev{discrimination, calibration, and composite metrics from predicted hospitalization probabilities $\hat{\pi}$. Discrimination reflects the model's ability to correctly distinguish hospitalized versus nonhospitalized patients based on lower versus higher $\hat{\pi}$, respectively. Calibration measures how well $\hat{\pi}$ aligned with observed hospitalization rates.}

\rev{For discrimination, we built receiver operating characteristic (ROC) curves and calculated the area under the ROC curve (AUC). Calibration was illustrated with plots of observed hospitalization rates versus average $\hat{\pi}$ by decile and numerically summarized by the calibration slope and intercept. As a hybrid assessment of model accuracy (discrimination and calibration together), we included Brier scores and sensitivity, specificity, positive predictive value (PPV), and negative predictive value (NPV) at the optimal threshold that maximized Youden's J statistic ($\text{sensitivity} + \text{specificity} - 1$). All numeric results are accompanied by $95\%$ confidence intervals ($95\%$ CIs). For the calibration slope and intercept, CIs were model-based; all others were bootstrapped. The Supplementary Materials contain details about these metrics and plots.}

We evaluated model performance in the full sample of $n = 707$ patients (optimistic) and via $5$-fold cross-validation (generalizable). \rev{We pooled predictions for out-of-fold model evaluations.} 
The missingness as a category models could not be cross-validated. Due to some extremely rare categories (like only $2$ patients missing BMI), not all levels were present when splitting the data into test and training folds. While we acknowledge the inflated metrics from the optimistic approach, they nonetheless shed light on the relative performance of the different missing data techniques and help pinpoint potential model overfitting. 

\section*{Results}

\subsection*{Cohort description}\label{res:cohort_desc}

The $707$ study participants had a median age of $49$ years old, and $59\%$ were female. Most were not hospitalized ($83\%$). \rev{The median follow-up time between earliest encounter anywhere and latest AHWFB encounter was $931$ days. Focusing on follow-up time in care at AHWFB, the median time between earliest and latest encounters during the study period was $52$ days, since these patients newly engaged in primary care. While most patients had years of total follow-up, it was possible that their first encounter at AWHFB was their only one before the study ended.} Additional summary statistics, including for the numeric ALI measurements, can be found in Table~\ref{tab:table1}. 

\begin{table}[h]
    \centering
    \begin{tabular}{lcc}
        \textbf{Variable} & \textbf{Non-Missing Values} & \textbf{Missing} \\
        \hline 
        \addlinespace
        \textit{Demographics} & & \\
        Age (Years) & $49$ ($33, 57$) & $0$ ($0\%$) \\
        Binary Biological Sex: Female & $418$ ($59\%$) & $0$ ($0\%$)\\
        Binary Biological Sex: Male & $289$ ($41\%$) & $0$ ($0\%$)\\
        \addlinespace
        \textit{ALI Components} & & \\
        Hemoglobin A1C & $5.7$ $(5.3, 6.8)$ & $403$ ($57\%$) \\
        Serum Albumin & $4.3$ $(4.1, 4.5)$ & $151$ ($21\%$) \\
        Body Mass Index & $29.0$ $(25.3, 33.9)$ & $2$ ($<1\%$) \\
        Cholesterol & $184.0$ $(161.0, 210.0)$ & $206$ ($29\%$) \\
        C-Reactive Protein & $3.9$ $(1.1, 55.2)$ & $701$ ($99\%$) \\
        Creatinine Clearance & $156.7$ $(156.7, 156.7)$ & $706$ ($>99\%$) \\
        Homocysteine & $10.2$ $(9.8, 12.2)$ & $701$ ($99\%$) \\
        Triglycerides & $119.0$ $(84.5, 176.3)$ & $206$ ($29\%$) \\
        Diastolic Blood Pressure & $77.0$ $(71.0, 82.2)$ & $0$ ($0\%$) \\
        Systolic Blood Pressure & $124.8$ $(117.0, 134.3)$ & $0$ ($0\%$) \\
        \addlinespace
        \textit{Clinical Encounters} &  & \\
        \addlinespace
        Total Follow-Up Time (Days) & $931$ $(570, 1112)$ & $0$ ($0\%$) \\
        Unique Encounter Dates & $12$ $(6, 21)$ & $0$ ($0\%$) \\
        AHWFB Follow-Up Time (Days) & $52$ $(0, 215)$ & $0$ ($0\%$) \\
        Hospitalized at AHWFB & $117$ ($17\%)$ & $0$ ($0\%$) \\
    \end{tabular}
    \caption{Summary of demographic characteristics and numeric allostatic load index (ALI) component measurements for the $707$ patients sampled from the electronic health records (EHR) at Atrium Health Wake Forest Baptist Hospital. Numeric variables are reported as median (quartile 1, quartile 3) and categorical variables as count ($\%$). Missingness is shown as count ($\%$). \rev{For the Clinical Encounters variables, data were filtered to include (i) encounters no more than $3$ years before a patient's first Atrium Health Wake Forest Baptist (AHWFB) primary care outpatient encounter and (ii) only encounters at least $1$ day prior to first hospitalization (for those who were hospitalized). Total Follow-Up Time was defined as the time between between patients' earliest encounter (at AHWFB or elsewhere) and latest AHWFB encounter before the end of the study period. The number of Unique Encounter Dates was taken from the same time period as Total Follow-Up Time, so it captured both AHWFB and elsewhere (from the lookback). AHWFB Follow-Up time was defined as the time between between patients' earliest and latest AHWFB encounters before the end of the study period.}} 
    \label{tab:table1}
\end{table}

The median complete-case proportion ALI was $0.33$, meaning that $50\%$ of patients \rev{had unhealthy values for $\leq 33\%$ of 
biomarkers} for which they had EHR data. After imputing with the best case (healthy) for all patients, this median decreased to $0.2$; if instead imputed with the worst case (unhealthy), it increased to $0.7$. As expected, the best and worst case scenario ALIs were less variable than the complete-case proportion (Supplemental Figure~S1). The median counts of unhealthy and missing ALI components were $2$ and $4$, respectively (Supplemental Figure~S2).

\subsection*{Missingness in ALI components}\label{subsec:res:miss.patterns}

\begin{figure}[h]
    \centering
    \includegraphics[width=\linewidth]{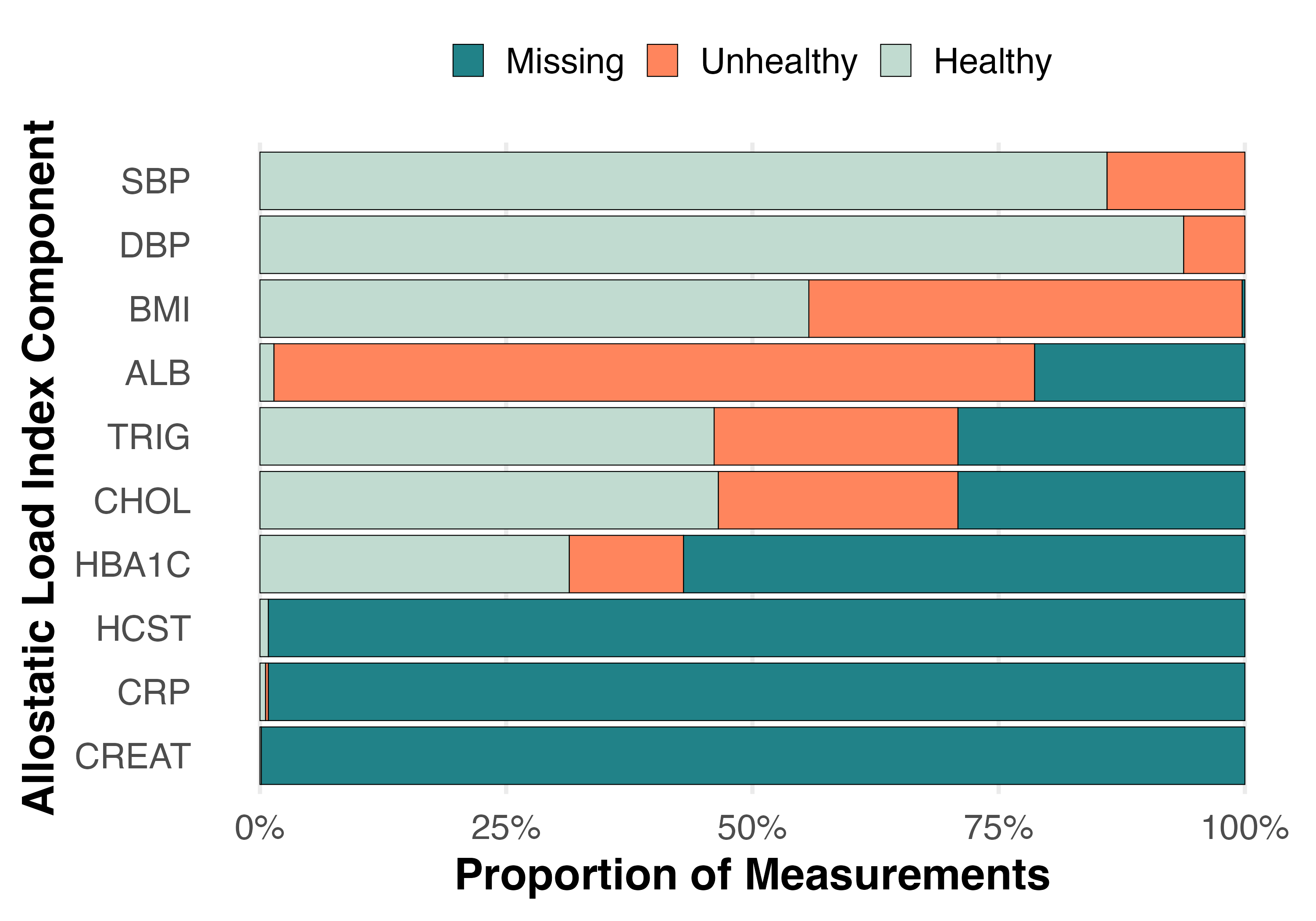}
    \caption{Proportion of $n = 707$ patients in the sample from the electronic health records EHR at Atrium Health Wake Forest Baptist Hospital with healthy, unhealthy, and missing values across the ten allostatic load index (ALI) components, after discretizing the original numeric biomarkers \rev{at their thresholds} from Table~\ref{tab:ali_components}. Abbreviations: hemoglobin A1C (HBA1C), serum albumin (ALB), body mass index (BMI), diastolic blood pressure (DBP), systolic blood pressure (SBP), cholesterol (CHOL), creatinine clearance (CREAT), C-reactive protein (CRP), homocysteine (HCST), and triglycerides (TRIG).}
    \label{fig:miss.barplot}
\end{figure}

EHR availability of ALI components depends on external factors, like whether a clinician deemed a particular test clinically necessary. 
No patients were missing SBP or DBP, and only two were missing BMI (Figure~\ref{fig:miss.barplot}). These vitals are expected to be very complete, because they are routinely collected during most encounters. 
In contrast, creatinine clearance, CRP, and homocysteine, which require specialized laboratory testing, demonstrated substantial missingness ($\geq 99\%$). 

\begin{figure}[h]
    \centering
    \includegraphics[width=\linewidth]{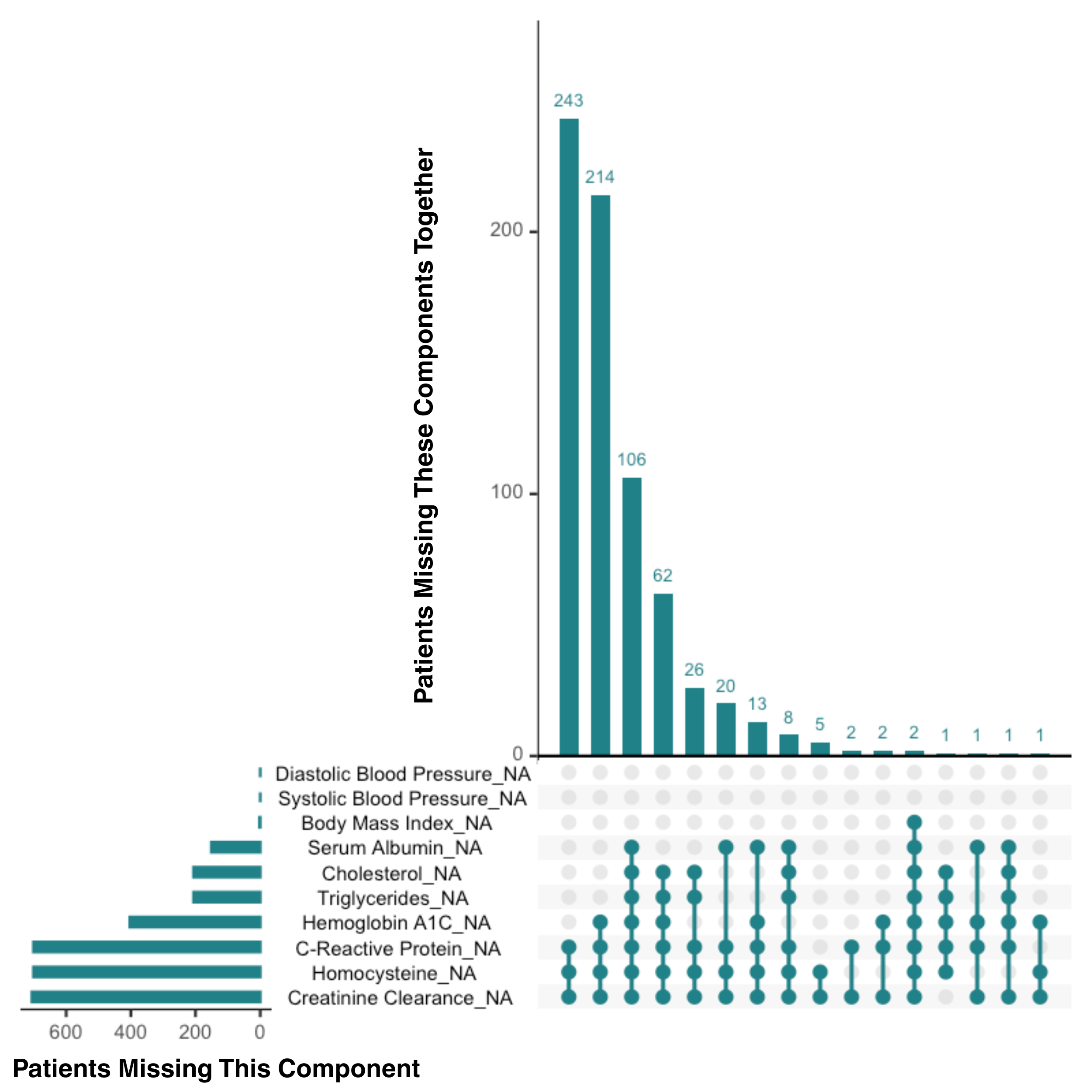}
    \caption{Upset plot displaying combinations of allostatic load index (ALI) components that were missing together for patients, made with the \textit{naniar} R package.\cite{naniar}}
    \label{fig:upset.plot}
\end{figure}

Across variables, missingness patterns could also reflect shared clinical ordering patterns (Figure~\ref{fig:upset.plot}). For example, homocysteine and CRP are typically ordered when further cardiovascular evaluation is warranted.\cite{Ridker2000,Dillon2010}  
Because these laboratory values are not part of routine screening panels, their absence likely reflects clinical judgment and makes their missingness informative.

\subsection*{Model comparisons}\label{subsec:res:modcomp}

\rev{Overall, discriminative performance was fair to low and the calibration low to moderate for all models considered. While some separate components methods had better discrimination than the summary measures in the full sample (higher AUC), their advantages did not persist when $5$-fold cross-validated  (especially for pattern submodels). For logistic regression, the summary measures offered much better calibration than the separate components. Calibration for random forests was more similar between these categories of missing data corrections. For both model types, the pattern submodels approach had substantial full-sample versus cross-validated performance differences, suggesting overfitting in the former. Brier scores were similar ($0.12$--$0.15$) and indicated good, but not great, overall accuracy for all models. 
Performance metrics for logistic regression and random forest models are summarized in Figures~\ref{fig:perform_logreg} and \ref{fig:perform_rf}, detailed in Supplemental Tables~S1 and S2, and discussed in the following subsections.} 

\subsubsection*{Best missing data approach for logistic regression}

\rev{Across summary measure models, AUC values were tightly clustered, falling within $0.01$ ($0.62$--$0.63$ full sample, $0.60$--$0.61$ cross-validated). 
While separate component methods seemed to offer better discrimination in the full sample (AUC $0.65$--$0.72$), their cross-validated performance was similar to summary measures. In particular, pattern submodels had the highest full-sample AUC but lowest cross-validated. 
} 



\rev{At the optimal thresholds, summary measure models tended to have better sensitivity ($\leq 0.79$), while separate component models had higher specificity ($\leq 0.73$, Supplemental Table~S1). 
All models had high NPV ($0.87$--$0.91$) but very low PPV ($\leq 0.30$). However, other thresholds (e.g., optimizing sensitivity) could improve these diagnostics. ROC curves can be found in Supplementary Figures~S3 and S4.}



\rev{
Summary measure models' predictions were fairly well calibrated (Supplemental Figures~S5). Separate component models’ average predictions were more spread out and better calibrated in the full sample than when cross-validated, with the most noticeable deviation for the highest average $\hat{\pi}$ decile (Supplemental Figure~S6). Calibration intercepts $\leq 0.03$ in magnitude suggested good calibration-in-the-large (mean calibration) for all cross-validated models. However, these models differed in their predictions' spread. With calibration slopes of $0.77$--$0.81$, summary measure models' predictions tended to be extreme (i.e., too low for nonhospitalized and too high for hospitalized). 
For pattern submodels and worst case imputation (separate), there was essentially no relationship between $Y$ and their predictions (calibration slopes $\approx 0$). Calibration slope and intercept were not calculated for full-sample logistic regression, since this model is mathematically guaranteed to (misleadingly) appear perfectly calibrated when fit and evaluated on the same data due to the underlying maximum likelihood estimation.\cite{Vach2013Calibration,VanCalster2016,Edlinger2022}}




\begin{figure}[h] 
    \centering
    \includegraphics[width=\linewidth]{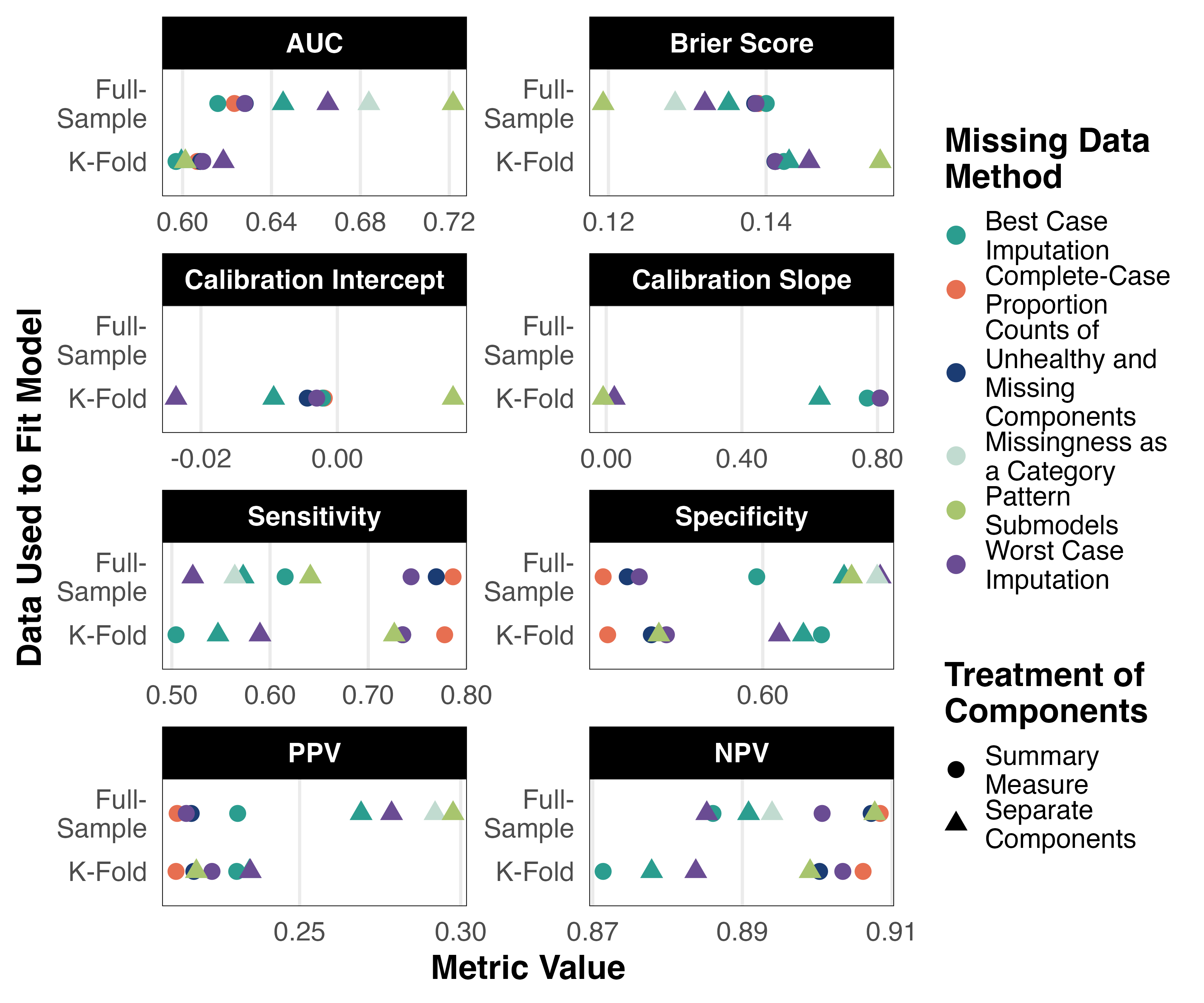}
    \caption{Various performance metrics for logistic regression using each missing data approach. 
    Due to very small counts for some of the separate components' categories, the missingness as a category approach could not be $5$-fold cross-validated. Calibration slope and intercept were not calculated for the full-sample logistic regression, since this model is almost guaranteed to appear perfectly calibrated when fit and evaluated on the same data.} 
    \label{fig:perform_logreg}
\end{figure}

\subsubsection*{Best missing data approach for random forest}


\rev{
Full-sample AUC values were also similar but slightly more spread out than for logistic regression, 
ranging from $0.56$ 
to $0.60$. 
Pattern submodels (AUC $=0.72$) was the only separate component model that seemed to offer improved discrimination over the summary measures in the full sample, while the others had AUC between $0.58$ and $0.61$. 
Again, out-of-sample performance for pattern submodels was not as good (AUC $= 0.60$), while other methods had AUC values within $0.02$ of their full-sample.  Overall, there was little meaningful difference in discriminative ability between summary measure and separate component random forest models.  

There was not a clear split between summary measure and separate component models in terms of whether sensitivity or specificity was favored at the optimal thresholds, and even the highest values were still moderate-to-low. 
All random forests exhibited very poor PPV ($\leq 0.30$) but good NPV ($0.85$--$0.91$). ROC curves can be found in Supplementary Figures~S7 and S8.


The complete-case proportion and best/worst case imputation (summary) appeared fairly well calibrated, but counts of unhealthy and missing components was not (Supplemental Figure~S9). Other than the full-sample pattern submodels (which appeared overfit), the separate component random forests were pretty poorly calibrated, particularly for the lowest and highest deciles (Supplemental Figure~S10). Still, all calibration intercepts were within $0.1$ of the target $0$, 
suggesting good calibration-in-the-large. 
The best and worst case (summary) imputation approaches had the closest calibration slopes to the ideal $1$; they declined when cross-validated (from $0.80$ to $0.62$ and $0.83$ to $0.64$, respectively) but remained closest. 
Predictions from the counts of unhealthy and missing components models were especially miscalibrated at the extremes (calibration slope $= 0.19$). Pattern submodels appeared perfectly calibrated in the full sample (calibration slope $= 1.04$) but its predictions unrelated to hospitalization when cross-validated (calibration slope = $-0.01$). Altogether, pattern submodels seemed to suffer from especially stark overfitting with the random forest, likely due to the model's complexity. 



}

\begin{figure}[h] 
    \centering
    \includegraphics[width=\linewidth]{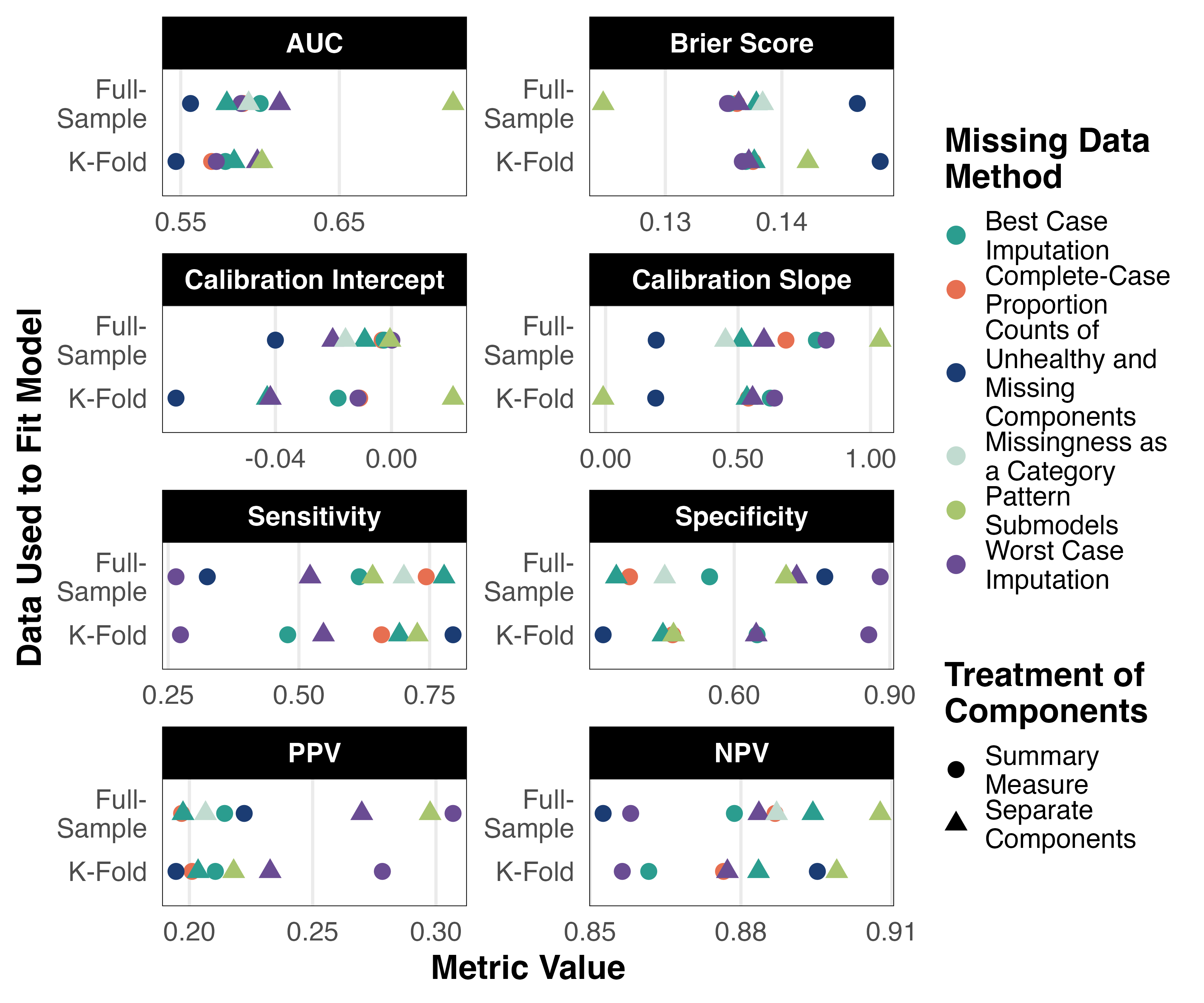}
    \caption{Various performance metrics for random forest using each missing data approach. Due to very small counts for some of the separate components' categories, the missingness as a category approach could not be $5$-fold cross-validated.}
    \label{fig:perform_rf}
\end{figure}

\section*{Discussion}\label{disc}

\rev{Our primary objective was pinpointing the missing data approach that yielded the best predictive performance for hospitalization from EHR-derived ALI when using logistic regression or random forest.} 
We assessed performance through two modeling lenses: (1) collapsing the ten components into a summary measure and (2) allowing each component to contribute separately to the models. In general, how we incorporated missing ALI components (or whether we did so at all) did not have much impact on prediction when using the summary measures but might if the components were included separately in the models. 

The summary measure models are appealing in practice. For example, these definitions could be embedded in electronic charts to give patients and clinicians a one-number summary to track overall health.\citep{Orkaby2024, Khurana2022, WholePersonHealth} Also, in past work, we and others have found these summary measures to have fairly strong \rev{inferential} associations with key clinical outcomes, like healthcare utilization.\cite{lotspeich2025} However, regardless of how missing ALI components were handled, the predictive performance using summary measures was poor to good. Looking forward, models allowing the ten ALI components to contribute separately can provide more personalized predictions and achieve better performance overall, and we hope that the missing data strategies discussed herein can provide a good starting place. However, larger samples will be needed.

Data-driven imputation,\cite{rubin2004multiple} predicting missing values from non-missing information, are more common than best/worst case in most applications. 
However, in EHR data, clinical decision-making often dictates whether labs are ordered for a particular patient\rev{. This} missingness is non-ignorable, since it is dictated by information we do not observe,\cite{Wells2013, Haneuse2021} and improperly modeling the missing data mechanism can actually harm analyses.\cite{Li2018} 

EHR data are inherently subject to other quality issues, like measurement error,\cite{Hersh2013, Kim2019, Nordo2019} which pose an important direction for future work. 
Prediction models could be strengthened by incorporating a partially validated subset (e.g., from expert chart review\cite{lotspeich2025}) to overcome measurement error in addition to missing data.\cite{Khudyakov2015} Further, the proposed missing data approaches could be applied to other deficit scores, like the EFI\citep{Pajewski2019} or index of cardiometabolic health (ICMH).\citep{Nobel2017} \rev{For the pattern submodels approach, we handled rare missing data patterns as proposed by Fletcher Mercaldo and Blume (2020).\cite{Fletcher2020} Methodological developments (e.g., with Bayesian partial pooling to borrow information between patterns) could improve upon this strategy.}

\rev{Herein, we hoped to identify which missing data approach to adopt after deciding on a particular prediction model. Our chosen models used only basic covariates and default settings. In future, we will build upon these findings to rigorously develop new prediction models using EHR-derived ALI. Tuning random forest hyper-parameters, considering other modeling methods, incorporating more  covariates, and comparing to ``benchmark'' models (like Epic's readmission prediction) will be instrumental to these developments, so that we can improve performance while keeping clinical relevance and deployment in mind.} 


\section*{Funding Statement}
The authors gratefully acknowledge Wake Forest University and Wake Forest University School of Medicine (under National Institutes of Health Grant UM1TR004929) for an intercampus collaborative grant that supported the data collection for this work.

\section*{Data Availability Statement}

Due to patient privacy, the authors do not have permission to share the EHR data. A simulated dataset for illustration is publicly available at \url{https://github.com/sarahlotspeich/missALI}, with \texttt{R} code for all models and missing data corrections discussed. 

\section*{Contributorship Statement}

Grayson Weavil (Conceptualization, Formal analysis, Methodology, Visualization, Writing - original draft), Joseph Rigdon (Conceptualization, Data curation, Funding acquisition, Writing - Review \& Editing), Sarah Lotspeich (Conceptualization, Data curation, Funding acquisition, Methodology, Software, Writing - original draft)

\section*{Conflict of Interests Statement}

The authors have no conflicts of interest to declare.

\bibliographystyle{vancouver}
\bibliography{sample}

\section*{Supplementary Materials}
\begin{itemize}
    \item \textbf{Additional details, tables, and figures:} The supplemental figures and tables referenced are available online at \url{https://github.com/sarahlotspeich/missALI/blob/main/Supplementary_Materials.pdf} as Supplementary Materials.
    \item \textbf{R code and data:} The R scripts and a simulated data to apply the methods discussed herein are available at \url{https://github.com/sarahlotspeich/missALI}.
\end{itemize}

\end{document}